\documentclass[twocolumn,floatfix,showpacs,prb,10pt]{revtex4-1}
\usepackage{graphicx}
\usepackage{amsmath}
\usepackage{amssymb}
\usepackage{bm}
\usepackage{hyperref}
\usepackage{multirow}
\usepackage{color}
\usepackage{amsbsy}
\usepackage{ulem}

\newcommand{\pd}{{\phantom\dag}}

\begin{document}

\title{Coupled-layer description of topological crystalline insulators}

\author{I. C. Fulga}
\author{N. Avraham}
\author{H. Beidenkopf}
\author{A. Stern}
\affiliation{Department of Condensed Matter Physics, Weizmann Institute of Science, Rehovot 76100, Israel}

\date{\today}
\begin{abstract}

We introduce a coupled-layer construction to describe three-dimensional topological crystalline insulators protected by reflection symmetry. Our approach uses stacks of weakly-coupled two-dimensional Chern insulators to produce topological crystalline insulators in one higher dimension, with tunable number and location of surface Dirac cones. As an application of our formalism, we turn to a simplified model of topological crystalline insulator SnTe, showing that its protected surface states can be described using the coupled-layer construction.

\end{abstract}
\maketitle

\section{Introduction}
\label{sec:intro}

The term \textit{topological insulator} (TI) was first coined to describe two- and three-dimensional systems that simultaneously host a gapped bulk and gapless boundary states protected by time-reversal symmetry (TRS).\cite{Kane2005a, Kane2005, Fu2007} The two-dimensional (2d) TI has one protected pair of helical modes on each edge, while the 3d TI shows one protected Dirac cone on each surface, which is pinned by time-reversal symmetry to one of the four time-reversal invariant momenta (TRIM) of the surface Brillouin zone (BZ). This peculiar structure of boundary modes apparently violates the fermion-doubling theorem, which states that in any system with TRS, Dirac cones must come in pairs.\cite{Nielsen1983} The resolution is that the second Dirac cone or helical edge mode pair appears on the opposite surface or edge, separated by the insulating bulk. This means that one cannot construct a lower-dimensional lattice model describing only a single boundary of a TI, nor neglect the presence of a topologically non-trivial bulk.

In contrast, some topological states of matter can, by construction, be thought of as structures formed out of weakly-coupled lower-dimensional building blocks. This is the case of weak topological insulators (WTI),\cite{Fu2007} formed by stacking many copies of a lower-dimensional TI. The first example of a WTI can be thought of as a layered 3d system consisting of many copies of a 2d TI stacked in the third dimension. It is adiabatically connected to its decoupled limit, meaning all WTI properties are recovered even for arbitrarily small inter-layer coupling. 
Due to the combination of TRS within each layer and translation symmetry along the stacking direction, WTI surfaces perpendicular to the layers show a pair of protected Dirac cones.

After the original WTI proposal, a large number of works have considered the interplay of topology and lattice symmetries.\cite{Slager2012, Jadaun2013, Chiu2013, Zhang2013, Benalcazar2014, Morimoto2013, Diez2015} It was discovered that not only translation, but also other lattice symmetries (rotation, reflection, glide symmetry, etc.) can lead to topologically non-trivial behavior. The latter phases were dubbed topological crystalline insulators (TCI),\cite{Fu2011, Hsieh2012} which may form both in 2d and 3d systems, and host gapless modes on boundaries preserving the protecting lattice symmetry. The experimental confirmation of a reflection symmetry protected TCI phase in 3d rocksalt crystal SnTe has sparked an intense activity in this field.\cite{Tanaka2012, Dziawa2012, Xu2012}

In this paper we introduce a coupled-layer model that can describe 3d TCIs protected by reflection symmetry. We analyze a system composed of weakly-coupled Chern insulators with alternating Chern number, extending the \textit{anti-ferromagnetic TI} (AFTI) model of Ref.~\onlinecite{Mong2010} to obtain TCIs protected by reflection symmetry. 
The coupled-layer approach can produce TCIs in which both the number and the location of protected surface modes can be tuned. The topological invariants of the layers determine the number of surface Dirac cones, the inter-layer coupling sets their position in the surface BZ, while the stacking direction controls which mirror symmetry is responsible for their protection. This provides an intuitive description which captures the properties of 3d TCIs protected by reflection symmetry.
Additionally, our construction enables us to examine the conditions under which 3d TCIs are adiabatically connected to the limit of decoupled layers, similar to Ref.~\onlinecite{Kim2015}, which studied a 3d TCI obtained by stacking 2d TCIs.

As a case study, we analyze the properties of rocksalt SnTe, a reflection symmetry protected TCI, in the coupled-layer framework. Using a simplified tight-binding model, we show that anisotropy can lead to a substantial increase in the number of surface Dirac cones: from four in the isotropic case to twelve with anisotropy. The protected surface modes can be understood in the language of coupled two-dimensional layers.

The rest of our work is organized as follows. In Section \ref{sec:AFTItoTCI}, we begin by reviewing the AFTI model (Section \ref{subsec:afti}) and show how it can be extended to produce TCIs with an arbitrary number of surface Dirac cones (Section \ref{subsec:tci}). We consider the effect of multiple non-trivial topological invariants, showing that the locations of Dirac cones in the surface BZ can be tuned using the coupled-layer construction (Section \ref{subsec:conepos}). Additionally, we examine how TCIs protected by multiple reflection symmetries can be modeled as systems of coupled layers (Section \ref{subsec:rxrz}). This approach can be used to understand the effect of anisotropy in 3d SnTe, as shown in Section \ref{sec:SnTe}. We conclude and discuss directions for future work in Section \ref{sec:conc}.

\section{Topological crystalline insulator from coupled Chern insulators}
\label{sec:AFTItoTCI}

In this Section we introduce a model of layered Chern insulators with alternating topological invariants, showing it describes the features of TCIs. To this end, we begin by summarizing the AFTI model of Ref.~\onlinecite{Mong2010} and its relation to strong TIs.

\subsection{Anti-ferromagnetic topological insulator}
\label{subsec:afti}

\begin{figure}[tb]
 \includegraphics[width=0.85\columnwidth]{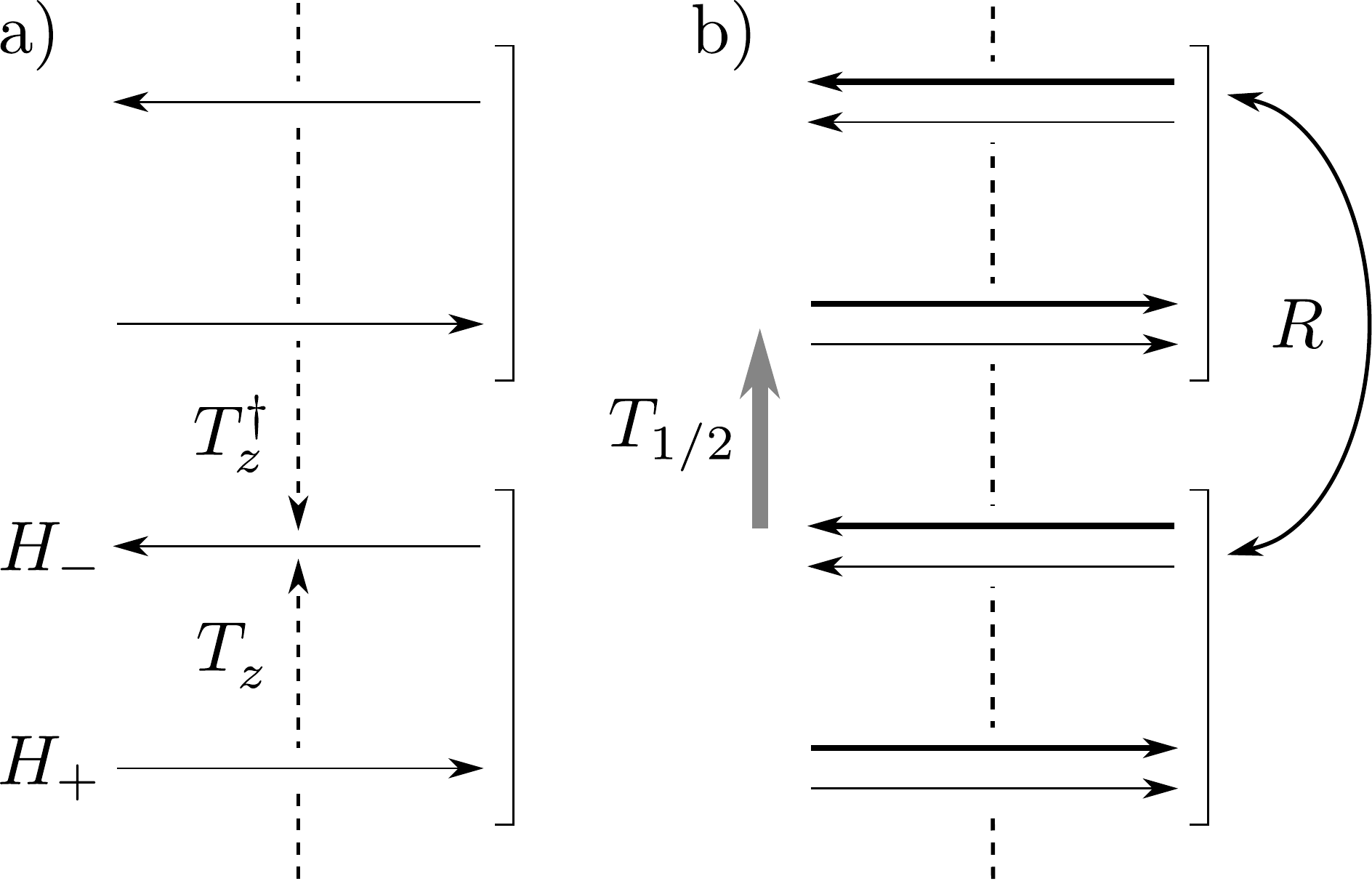}
 \caption{Weakly coupled Chern insulating layers, $H_{\pm}$, stacked in the $z$-direction, with invariants $C_\pm=\pm1$ (panel a), or $C_\pm=\pm2$ (panel b). The unit cell (square bracket) is composed of two layers, each of which carries the same number of chiral modes (horizontal arrows). The hopping matrix $T_z$ acts both within a unit cell and between neighboring ones. In both cases, the systems preserve the ${\cal S} = \Theta T_{1/2}$ symmetry: $\Theta$ reverses edge mode chirality, interchanging $H_+$ and $H_-$, while $T_{1/2}$ translates the system by one layer. This leads to equal coupling between chiral edge modes and a gapless surface in (a), but allows for a fully gapped surface in (b), obtained by coupling different pairs of modes: thick arrow in one layer to thin arrow in the layer above. Imposing reflection symmetry $R$ leads to gapless surfaces in both cases, protected by a $\mathbb{Z}$ valued mirror Chern number $C_{M}$.\label{fig:afti}}
\end{figure}

As mentioned, a 3d strong TI has one Dirac cone on any surface, regardless of orientation, which is protected by time-reversal symmetry. Due to this property, the system cannot be fully described in the language of weakly coupled, topologically non-trivial layers.
It was shown however that some of the features of a 3d TI surface can be mimicked by a stacked structure consisting of non-trivial layers with alternating Chern number $\pm1$ (see Fig.~\ref{fig:afti}a). The resulting model, called an anti-ferromagnetic topological insulator,\cite{Mong2010, Baireuther2014} hosts a single protected Dirac cone on surfaces parallel to the stacking direction. Similar to the 3d TI, the AFTI is characterized by a quantized magnetoelectric effect, leading to a term in the electromagnetic Lagrangian of the form\cite{Mong2010}
\begin{equation}\label{eq:thetaterm}
\Delta{\cal L}_{\rm EM} = \frac{\theta e^2}{2\pi h}\vec{E}\cdot\vec{B},
\end{equation}
with a quantized value of $\theta=\pi$.
The comparison between the 3d TI and the AFTI is only approximate however, as in a strong 3d TI all surfaces have a single Dirac cone, whereas in the AFTI some surfaces are gapped.

Due to the alternating sign of Chern numbers $\pm1$ in adjacent layers of the AFTI, the chiral edge mode present in each layer may couple with a counter-propagating neighboring one and gap out. However, due to translation symmetry the coupled system has protected surface modes, which can be described in terms of a $\mathbb{Z}_2$ topological classification. The protecting symmetry is anti-unitary, ${\cal S} = \Theta T_{1/2}$, with $\Theta$ the anti-unitary time-reversal operator ($\Theta^2=-1$) and $T_{1/2}$ translation by one layer, \textit{i.e.}~half of the unit cell. If a surface preserves the combined symmetry ${\cal S}$, then no dimerization between counter-propagating edge modes is possible, rendering the entire surface gapless.

Consider an AFTI stacked along the $z$-direction, with Hamiltonian
\begin{equation}\label{eq:afti_stack}
 {\cal H}({\bf k}) = \begin{pmatrix}
                 H_+(k_x,k_y) & T_z^\dag + T_z e^{i k_z} \\
                 T_z + T_z^\dag e^{-ik_z} & H_-(k_x, k_y)
                \end{pmatrix},
\end{equation}
where $H_{\pm}$ are the Hamiltonians of adjacent layers (Chern numbers $C_\pm = \pm1$), and $T_z$ the inter-layer hopping matrix, which connects layers both within the same unit cell, as well as between neighboring cells (see Fig.~\ref{fig:afti}a). Imposing that adjacent layers are time-reversed partners,
\begin{equation}\label{eq:trs_plusmin}
 \Theta H_\pm(k_x,k_y) \Theta^{-1}=H_\mp(-k_x,-k_y),
\end{equation}
enables to define a translation operator by one layer,
\begin{equation}\label{eq:t12}
 T_{1/2}=\begin{pmatrix}
          0 & 1 \\
          e^{-ik_z} & 0
         \end{pmatrix},
\end{equation}
such that ${\cal S}=\Theta T_{1/2}$ is a momentum dependent anti-unitary operator which commutes with the Hamiltonian: ${\cal S} {\cal H}({\bf k}) {\cal S}^{-1} = {\cal H}({\bf -k})$ if $T_z$ is time-reversal symmetric. On the $k_z=0$ plane, since $\Theta^2=-1$, the combined symmetry ${\cal S}^2 = -1$ effectively becomes a two-dimensional time-reversal symmetry, enabling a $\mathbb{Z}_2$ topological classification of the system similar to the 2d TI, the quantum spin Hall effect. For layers which are finite in the $x$-direction, a single protected surface Dirac cone forms in the $y-z$ surface BZ, which is pinned to a TRIM point.

Owing to the $\mathbb{Z}_2$ structure of its invariant, the AFTI is non-trivial whenever the Chern numbers of $H_{\pm}$ are odd, and trivial when they are even. Indeed, when the layers carry alternating invariants $C_+=+2$ and $C_-=-2$, it is possible to fully gap out the surface without breaking the ${\cal S}$ symmetry. This can be achieved by coupling different pairs of chiral modes, as shown in Fig.~\ref{fig:afti}b. Choosing a hopping matrix that connects the top mode of each layer (thick arrow) to the bottom mode of the layer above (thin arrow) allows for a pairwise gapping of chiral edge states.

\subsection{Topological crystalline insulator}
\label{subsec:tci}

Our aim is to extend the model \eqref{eq:afti_stack} to describe three-dimensional topological crystalline insulators. To this end, rather than focusing on the effect of ${\cal S}$, we define a new protecting symmetry: reflection about one layer. The Hamiltonian \eqref{eq:afti_stack} now obeys the constraint
\begin{equation}\label{eq:refl_sym}
 R(k_z) {\cal H}(k_x,k_y,k_z) R(k_z)^{-1} = {\cal H}(k_x,k_y,-k_z),
\end{equation}
where $R(k_z)$ is a unitary, momentum-dependent reflection operator.

To prevent the possibility of a gapped surface when $C_\pm$ are even, we choose a reflection operator that does not mix different edge modes within a layer.
If $R(k_z)$ acts trivially on the degrees of freedom associated with the diagonal blocks of the Hamiltonian \eqref{eq:afti_stack}, $H_+$ and $H_-$, then a pairwise gapping of edge modes is no longer allowed. For any hopping matrix of the form shown in Fig.~\eqref{fig:afti}b, connecting different edge modes in one layer and the layer above (thick arrow to thin arrow), the constraint \eqref{eq:refl_sym} forces the existence of a coupling of the same type to the layer below. This produces a gapless surface regardless of the parity of $C_\pm$, turning the layered system ${\cal H}$ into a model for a 3d TCI.

Due to reflection symmetry, any number of surface Dirac cones is allowed. Their number is given by the number of chiral modes present in each layer, $H_\pm$, and their location in the surface BZ is controlled by the structure of the inter-layer coupling matrix. If $R(k_z)$ does not rotate the degrees of freedom associated to $H_\pm$, the relation \eqref{eq:refl_sym} also constrains the form of $T_z$, which can be either a hermitian matrix, or an anti-hermitian one.

We examine first the case in which the inter-layer coupling is hermitian, $T^\pd_z=T^\dag_z$. The upper off-diagonal block of \eqref{eq:afti_stack} now takes the form $T_z (1+e^{ik_z})$, such that the Chern insulating layers $H_\pm$ decouple at $k_z=\pi$. The corresponding reflection operator reads\cite{Fulga2014}

\begin{equation}\label{eq:refl_pi}
 R(k_z)=\begin{pmatrix}
    1 & 0 \\
    0 & e^{ik_z}
   \end{pmatrix}.
\end{equation}

For layers that are finite in the $x$-direction, the $y-z$ surface BZ now shows a total of $|C_+|$ pairs of counter-propagating modes pinned to the $k_z=\pi$ line, which occur at opposite $k_y$ momenta due to the TRS constraint \eqref{eq:trs_plusmin}. Away from the reflection symmetric $k_z=\pi$ line, the Hamiltonian \eqref{eq:afti_stack} is no longer block-diagonal, enabling the counter-propagating modes to couple and gap out. As such, the surface BZ shows a total of $|C_+|$ surface Dirac cones, whose Dirac points are pinned to $k_z=\pi$, and which are protected by reflection symmetry.

The topological classification of the AFTI in terms of the anti-unitary ${\cal S}$ symmetry had a $\mathbb{Z}_2$ structure, such that only systems with an odd Chern number per layer had protected surface states. In contrast, when the protection is expressed in terms of the unitary reflection operator $R(k_z)$, a $\mathbb{Z}$ classification emerges, where any number of surface Dirac cones are allowed.
By imposing the reflection symmetry \eqref{eq:refl_pi} on the stack of Chern insulators we have obtained a model for a three-dimensional TCI. No assumptions have been made on the magnitude of inter-layer coupling $T_z$, so it can be made arbitrarily weak. As such, the Hamiltonian \eqref{eq:afti_stack} is adiabatically connected to its decoupled limit, similar to WTIs, but has the properties of a TCI phase.

For instance, the gapless surface is protected by mirror, as opposed to time-reversal symmetry between neighboring layers.\cite{Hsieh2012}
To see this, consider adding a small perturbation to the $H_+$ layers, which does not change the number of edge modes but violates the TRS constraint \eqref{eq:trs_plusmin}. If the mirror symmetry of Eqs.~\eqref{eq:refl_sym} and \eqref{eq:refl_pi} is preserved, the surface remains gapless due to the decoupling of $H_+$ and $H_-$ sectors at $k_z=\pi$.
As long as neighboring layers are time-reversed partners however, their Chern numbers must be opposite, $C_+=-C_-$, such that the surface hosts $|C_\pm|$ Dirac cones. Therefore, while reflection symmetry is required to produce a gapless surface, time-reversal symmetry is not a necessary condition.

We characterize the 3d system in terms of the topological invariant associated with 3d TCIs: the mirror Chern number. The latter is defined as $C_{k_z,\pi} = (C_+-C_-)/2$, where the subscript denotes the location of the mirror invariant plane ($k_z=\pi$). There are a total of $|C_{k_z,\pi}|$ protected Dirac cones on any surface preserving reflection symmetry, which are formed by the gapless modes of $H_+$ and $H_-$, and which occur at opposite momenta due to the TRS constraint \eqref{eq:trs_plusmin}. When $C_{k_z,\pi}$ is odd, some Dirac cones will be pinned to TRIM points of the surface BZ. An even value of $C_{k_z,\pi}$ allows for surface Dirac cones which are no longer pinned to TRIM points, but can slide in the surface BZ along the projection of the mirror plane. This reproduces the features of surface states in SnTe,\cite{Liu2013a} as well as those predicted in TCI proposals based on super-lattices.\cite{Yang2014, Li2014}

\subsection{Tunable Dirac cone positions}
\label{subsec:conepos}

The layered system \eqref{eq:afti_stack} provides a simple way of constructing 3d TCI models with an arbitrary number of surface Dirac cones. Furthermore, the location of Dirac cones can be tuned by modifying the inter-layer coupling matrix. For a hermitian $T^\pd_z=T^\dag_z$, protected modes occur at $k_z=\pi$ in the surface BZ. If instead we choose an anti-hermitian coupling, $T^\pd_z=-T_z^\dag$, the upper off-diagonal block of \eqref{eq:afti_stack} now reads $T_z(e^{ik_z}-1)$, such that it becomes block-diagonal at $k_z=0$. The same arguments apply, but now with a reflection operator which takes a different form,
\begin{equation}\label{eq:refl_zero}
  R(k_z)=\begin{pmatrix}
    1 & 0 \\
    0 & -e^{ik_z}
   \end{pmatrix},
\end{equation}
and with a mirror Chern number that counts protected Dirac cones on the $k_z=0$ line of the surface BZ: $C_{k_z,0} = (C_+-C_-)/2$.

Depending on the space group of the BZ, a single reflection symmetry constraint of the form \eqref{eq:refl_sym} may allow for either one or two mirror-symmetric planes in the BZ. In the following we focus on the case in which two mirror invariant planes exist, while in Section \ref{sec:SnTe} we will examine a model in which there is only one such plane associated to a reflection symmetry.

Consider a system in which a $z\to-z$ mirror symmetry leads to two mirror planes in the BZ, $k_z=0,\pi$, and two associated mirror Chern numbers: $C_{k_z,0}$ and $C_{k_z,\pi}$. The importance of considering both topological invariants was first stressed in Ref.~\onlinecite{Kim2015}, which studied a layered 3d TCI formed not by stacking Chern insulators, but 2d TCI layers. In this construction the weakly coupled limit has $C_{k_z,0}=C_{k_z,\pi}$, showing an equal number of surface Dirac cones on both lines of the surface BZ. In contrast, our scheme allows to address each of the two mirror Chern numbers separately, allowing for 3d TCI models with different numbers of Dirac cones on different mirror-symmetric lines.

To obtain a model with arbitrary values of $C_{k_z,0}$ and $C_{k_z,\pi}$, we construct a combined Hamiltonian with an increased number of orbitals per unit cell.\cite{Teo2010, Diez2015} One set of orbitals describes a stack of Chern insulators with a hermitian hopping matrix, $T^\pd_z=T^\dag_z$, while the other set has $T^\pd_z=-T^\dag_z$. The combined Hamiltonian reads
\begin{equation}\label{eq:h_combined}
 H = {\cal H}_0 \oplus {\cal H}_\pi \equiv \begin{pmatrix}
                                     {\cal H}_0 & \Lambda \\
                                     \Lambda^\dag & {\cal H}_\pi
                                    \end{pmatrix},
\end{equation}
with ${\cal H}_{0,\pi}$ of the form \eqref{eq:afti_stack}, and $\Lambda$ a coupling matrix which does not close the bulk gap and preserves the full set of symmetries of each sub-block. Since Chern numbers form an additive group,\cite{Teo2010,R2010,Wen2012} tuning the number of chiral modes in the layers of ${\cal H}_{0,\pi}$ allows for 3d TCI models with arbitrary numbers of surface Dirac cones, both at $k_z=0$ and at $k_z=\pi$. The parity of the number of surface Dirac cones determines the appearance of a topological $\theta$-term of the form Eq.~\eqref{eq:thetaterm}, like in 3d strong TIs.\cite{Fang2012, Varjas2015} When the BZ has a single mirror-invariant plane associated to a reflection symmetry (for instance $k_z=0$), then $\theta = \pi C_{k_z,0} \mod 2\pi$, while in the case of two mirror planes $\theta=\pi(C_{k_z,0} - C_{k_z,\pi})\mod 2\pi$.

\begin{figure}[t!]
 \includegraphics[width=0.5\columnwidth]{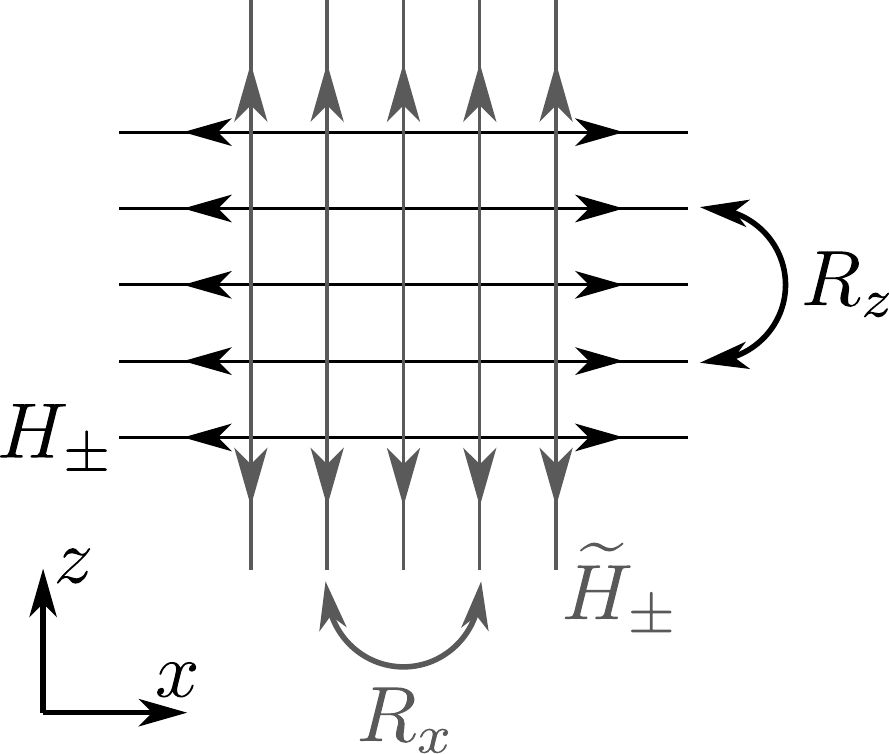}
 \caption{System composed of Chern insulators with alternating Chern numbers, stacked both in the $x$ and in the $z$ directions, shown as gray and black arrows, respectively. The pairs of layers stacked in $z$, with Hamiltonians $H_+$ and $H_-$, must lie on top of each other in order to preserve the $R_x$ reflection symmetry. The same applies to the layers stacked in the $x$ direction, denoted by $\widetilde{H}_+$ and $\widetilde{H}_-$. \label{fig:stackxz}}
\end{figure}

Note that the combined Hamiltonian \eqref{eq:h_combined} remains adiabatically connected to its decoupled limit, since both of its sub-blocks are protected by a mirror symmetry acting in the $z$-direction. The magnitudes of inter-layer couplings entering in ${\cal H}_{0,\pi}$, as well as the magnitude of $\Lambda$ can be chosen to be arbitrarily weak, while still preserving the full set of TCI features. This property however is lost when considering the effect of multiple reflection symmetries.

\subsection{Multiple reflection symmetries}
\label{subsec:rxrz}

Surface Dirac cones protected by different reflection symmetries can be obtained by combining Hamiltonians ${\cal H}_i$ describing Chern insulating layers stacked in different directions.
Consider for instance a Hamiltonian of the form Eq.~\eqref{eq:h_combined}, where the sub-blocks describe Chern insulators layered in the $z$ and in the $x$ directions, as shown in Fig.~\ref{fig:stackxz}. Unlike the setup of Fig.~\ref{fig:afti}, in order for the full Hamiltonian to respect both the $x\to -x$ and $z\to -z$ mirror symmetries,

\begin{equation}\label{eq:refl_x}
 R^\pd_x H(k_x,k_y,k_z) R_x^{-1} = H(-k_x,k_y,k_z),
\end{equation}
\begin{equation}\label{eq:refl_z}
 R^\pd_z H(k_x,k_y,k_z) R_z^{-1} = H(k_x,k_y,-k_z),
\end{equation}
pairs of layers with opposite Chern numbers must lie on top of each other, such that they are mapped into each other by the reflection operators $R_x$ or $R_z$. As such, for the layers stacked in the $z$ direction with Hamiltonians $H_+$ and $H_-$, we impose
\begin{equation}\label{eq:layer_reflx}
 H_+(k_x,k_y) = H_-(-k_x,k_y)
\end{equation}
rather than the TRS constraint of Eq.~\eqref{eq:trs_plusmin}. Note that Eq.~\eqref{eq:layer_reflx} also guarantees that $H_+$ and $H_-$ have opposite Chern numbers. Similarly, for the Chern insulating layers stacked in the $x$ direction,
\begin{equation}\label{eq:layer_reflz}
 \widetilde{H}_+(k_z,k_y) = \widetilde{H}_-(-k_z,k_y).
\end{equation}

If each of the mirror symmetry relations Eq.~\eqref{eq:refl_x} and \eqref{eq:refl_z} allow for two mirror invariant planes, $k_{x,z}=0$ and $\pi$, the full system is characterized by four mirror Chern numbers: $C_{k_x,0}$, $C_{k_x,\pi}$, $C_{k_z,0}$, and $C_{k_z,\pi}$. In general, the simultaneous presence of multiple symmetries can constrain the allowed values of topological invariants.\cite{Fang2012, Alexandradinata2014, Dong2016, Varjas2016} For the system considered in Fig.~\ref{fig:stackxz}, it was shown that if the two reflection operators commute, $[R_x,R_z]=0$ like for spinless fermions or bosons, all four mirror Chern numbers must vanish,\cite{Alexandradinata2014, Dong2016} such that only topologically trivial systems are possible. Further analysis shows that for anti-commuting reflection operators, $\{ R_x, R_y \}=0$, a gapped bulk spectrum requires the sum of the four mirror Chern numbers to be even.\cite{Alexandradinata2014, Dong2016}

We choose a block-diagonal Hamiltonian of the form
\begin{widetext}
\begin{equation}\label{eq:hrxrz}
 H = \begin{pmatrix}
      H_+(k_x,k_y) & T_z \sin(k_z) & 0  & 0  \\
      T_z \sin(k_z) & H_-(k_x,k_y) & 0  & 0  \\
      0  & 0  & \widetilde{H}_+(k_z,k_y) & T_x \sin(k_x) \\
      0  & 0  & T_x \sin(k_x) & \widetilde{H}_-(k_z,k_y)
     \end{pmatrix},
\end{equation}
\end{widetext}
which, given Eqs.~\eqref{eq:layer_reflx} and \eqref{eq:layer_reflz}, obeys the mirror symmetries \eqref{eq:refl_x} and \eqref{eq:refl_z} with anti-commuting reflection operators
\begin{equation}\label{eq:rxmat}
 R_x = \begin{pmatrix}
        0 & 1 & 0 & 0 \\
        1 & 0 & 0 & 0 \\
        0 & 0 & 1 & 0 \\
        0 & 0 & 0 & -1 \\
       \end{pmatrix},
\end{equation}
and
\begin{equation}\label{eq:rzmat}
 R_z = \begin{pmatrix}
        1 & 0 & 0 & 0 \\
        0 & -1 & 0 & 0 \\
        0 & 0 & 0 & 1 \\
        0 & 0 & 1 & 0 \\
       \end{pmatrix}.
\end{equation}

As before, at either $k_z=0,\pi$ or $k_x=0,\pi$ the Chern insulating layers decouple, enabling the formation of surface Dirac cones. The full system has mirror Chern numbers which are pairwise equal, with $C_{k_z,0}=C_{k_z,\pi}$ given by the number of chiral edge modes of the $H_+$ and $H_-$ layers, while $C_{k_x,0}=C_{k_x,\pi}$ are determined by the number of edge modes of $\widetilde{H}_+$ and $\widetilde{H}_-$. 
Unequal topological invariants may be obtained by folding the surface BZ along one momentum direction without breaking the mirror symmetries. Adding longer ranged hopping terms in the $x$ direction, for instance to the $H_\pm$ layers, will lead to a change of mirror Chern numbers $C_{k_x,0} \to 2 C_{k_x,0}$ and $C_{k_x,\pi} \to 0$.
In Appendix A we study a concrete tight-binding model realizing a system with two mirror symmetries, and discuss other methods to obtain unequal invariants $C_{k_x,0}\neq C_{k_x,\pi}$, based on inter-layer couplings which are not arbitrarily weak.

For a system which is finite in the $y$ direction, the $x - z$ surface BZ will show two sets of Dirac cones, protected by the $R_x$ and $R_z$ mirror symmetries, respectively.
When considered separately, each set of Dirac cones already appears in the weakly-coupled limit. Together however, they form a system that is strongly coupled in all three directions, and can no longer be thought of as almost decoupled two-dimensional layers. For instance, reducing both the inter-layer coupling of $H_\pm$, $T_z$, as well as the $z$ direction hopping of $\widetilde{H}_\pm$, would eventually cause the latter to undergo a bulk gap closing.

The power of the coupled-layer framework is that it provides an intuitive way of generating 3d TCIs protected by reflection symmetry. The number of surface Dirac cones can be traced back to the number of chiral modes in each layer, while their location in the surface BZ can be selected by tuning the stacking directions and inter-layer couplings. One of the drawbacks is that modeling the effect of multiple reflection symmetries, for instance by means of Eq.~\eqref{eq:hrxrz}, rapidly increases the number of orbitals per unit cell, which can exceed that of other TCI tight-binding models.

\section{Anisotropic tin telluride}
\label{sec:SnTe}

In this Section, we study a tight-binding model of topological crystalline insulator SnTe, showing that its protected boundary states can be interpreted in terms of the coupled-layer construction. Irrespective of the number of surface Dirac cones, each reflection symmetry enables us to rotate the Hamiltonian to the form of Eq.~\eqref{eq:afti_stack}, corresponding to a stack of Chern insulators with alternating topological invariants. On the mirror symmetric lines of the surface BZ the layers become decoupled, leading to the formation of surface Dirac cones.

In SnTe, the bands near the Fermi level are mainly composed of the $p$-orbitals of the two atomic species. Motivated by this fact, we use a simplified tight-binding model constructed from the Wannier functions corresponding to the three $p$-orbitals of Sn and Te atoms.\cite{Hsieh2012} The real space Hamiltonian reads
\begin{equation}\label{eq:HSnTe}
\begin{split}
H = & m \sum_j (-1)^j \sum_{{\bf r},s} {\bf c}^\dag_{js}({\bf r})\cdot{\bf c}^\pd_{js}({\bf r}) + \\
& \sum_{j,j'} t^\pd_{j,j'} \sum_{({r},{r'}),s} {\bf c}^\dag_{js}({\bf r})\cdot {\bf d}^\pd_{\bf rr'}\, {\bf d}^\pd_{\bf rr'}\cdot {\bf c}^\pd_{j's}({\bf r'})+h.c. \\
& + \sum_j i\lambda_j \sum_{{\bf r},s,s'} {\bf c}^\dag_{js}({\bf r})\times{\bf c}^\pd_{js'}({\bf r})\cdot{\boldsymbol\sigma}_{ss'},
\end{split}
\end{equation}
where $j=1,2$ labels the atomic species (Sn or Te), $s=\uparrow, \downarrow$ labels the electron spin, and ${\bf r}$ is the site position in the cubic lattice. The creation and annihilation operators ${\bf c}^\dag$ and ${\bf c}^\pd$ are three component vectors, corresponding to the three $p$-orbitals of Sn and Te, with $m$ the potential difference between them. The unit vectors ${\bf d}_{\bf rr'}$ point from lattice site ${\bf r}$ to ${\bf r'}$, so that $t_{12}=t_{21}$ describe nearest neighbor hopping from Sn to Te, while $t_{11}$ and $t_{22}$ are next-nearest neighbor couplings within the same sublattice (Sn to Sn, or Te to Te). As such, the second term in Eq.~\eqref{eq:HSnTe} models $\sigma$-bond hopping between the $p$-orbitals. The term $\lambda_j$ is the strength of atomic spin-orbit coupling, of the form ${\bf L}\cdot\boldsymbol\sigma$, with $\boldsymbol\sigma$ the vector of Pauli matrices in spin space, and ${\bf L}$ the orbital angular momentum of the $p$-orbitals in the cartesian basis:
\begin{equation}\label{eq:lmatrices}
\begin{split}
L_x = \begin{pmatrix}
        0 & 0 & 0 \\
        0 & 0 & -i \\
        0 & i & 0
      \end{pmatrix}, \\
L_y = \begin{pmatrix}
        0 & 0 & i \\
        0 & 0 & 0 \\
        -i & 0 & 0
      \end{pmatrix}, \\
L_z = \begin{pmatrix}
        0 & -i & 0 \\
        i & 0 & 0 \\
        0 & 0 & 0
      \end{pmatrix}.
\end{split}
\end{equation}

\begin{figure}[tb]
 \includegraphics[width=0.7\columnwidth]{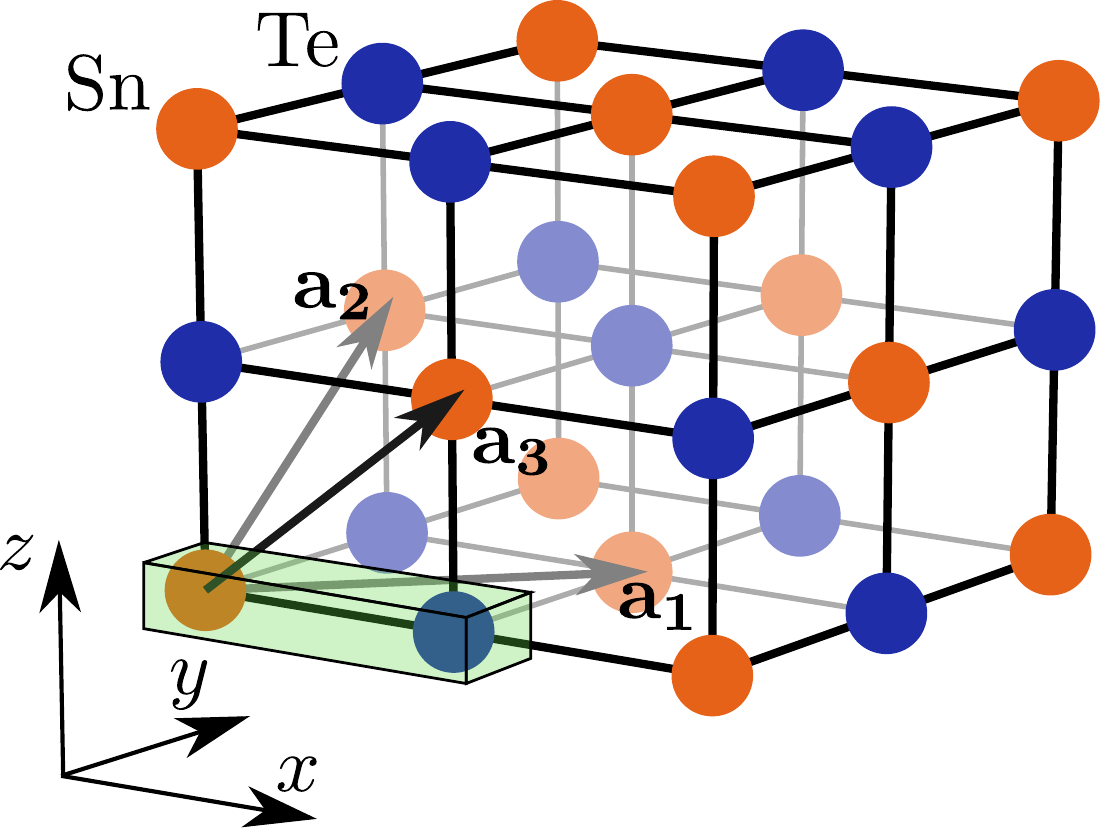}
 \caption{Face-centered cubic structure of rocksalt SnTe. The unit cell (green box) is composed of one Sn (orange) and one Te (dark blue) atom. The principal crystal directions, $x$, $y$, and $z$, as well as the Bravais vectors ${\bf a_1}$, ${\bf a_2}$, and ${\bf a_3}$ are indicated by arrows.\label{fig:fcc}}
\end{figure}

The Hamiltonian \eqref{eq:HSnTe} describes a face-centered cubic lattice with two atoms per unit cell, located at ${\bf r_1} = 0$ and ${\bf r_2}= a/2 \,{\hat x}$, with $a$ the lattice constant and ${\hat x}$, ${\hat y}$, ${\hat z}$ unit vectors pointing along the three principal directions, as shown in Fig.~\ref{fig:fcc}. We choose the Bravais vectors ${\bf a_1} = a/2 ({\hat x} + {\hat y})$, ${\bf a_2} = a/2 ({\hat y} + {\hat z})$, and ${\bf a_3} = a/2 ({\hat z} + {\hat x})$, leading to reciprocal lattice vectors ${\bf k_1} = 2\pi/a ({\hat x} + {\hat y} - {\hat z})$, ${\bf k_2} = 2\pi/a (-{\hat x} + {\hat y} + {\hat z})$, and ${\bf k_3} = 2\pi/a ({\hat x} - {\hat y} + {\hat z})$.
The tight-binding model has multiple lattice symmetries, including $C_4$ rotation symmetries around the lattice sites and mirror symmetries along all principal axes. Additionally, it shows reflection symmetries with respect to 6 equivalent $(110)$ mirror planes,\cite{Hsieh2012} each of which has a single mirror invariant plane in the 3d BZ.

We choose $m=1.65$, $t_{12}=t_{21}=0.9$, $t_{11}=-t_{22}=0.5$, and $\lambda_1=\lambda_2=0.7$.\cite{talkHsieh} For these parameters, the model \eqref{eq:HSnTe} reproduces the main features of 3d topological crystalline insulator SnTe. In an infinite slab geometry with boundaries along the $z$-direction, Miller indices (001), we find four surface Dirac cones, two lying on the projection of the (110) plane of the surface BZ, and two on the (1$\overline{1}$0) plane. The topological nature of the surface states can be determined by computing the mirror Chern numbers of bulk SnTe along these two planes. The momentum space form of Hamiltonian \eqref{eq:HSnTe} obeys a (110) reflection symmetry mapping $x\to y$ and $y\to x$,
\begin{equation}\label{eq:SnTe110}
 M_{xy}^\dag H(k_1, k_2, k_3) M_{xy}^\pd = H(k_1, k_3, k_2).
\end{equation}

On the (110) plane, $k_2=k_3$ and the mirror operator becomes momentum-independent. It is diagonal in the degree of freedom associated to the atomic species, since it maps Sn to Sn and Te to Te:
\begin{equation}\label{eq:Mxy}
 M_{xy} = \begin{pmatrix}
           {\cal M}_{xy} & 0 \\
           0 & {\cal M}_{xy}
          \end{pmatrix}.
\end{equation}

For each one of the two atoms in the unit cell, mirror symmetry involves both an interchange of the the $p_x$ and $p_y$ orbitals, as well as a spin rotation:
\begin{equation}\label{eq:smallMxy}
 {\cal M}_{xy}    = \begin{pmatrix}
        0 & 1 & 0 \\
        1 & 0 & 0 \\
        0 & 0 & 1
       \end{pmatrix}
\otimes \frac{i(\sigma_x - \sigma_y)}{\sqrt{2}}.
\end{equation}

Due to the spin component of \eqref{eq:smallMxy}, applying the reflection operator twice amounts to a $2\pi$ rotation of spin, such that $M_{xy}^2=-1$. The mirror eigenvalues are therefore $m=\pm i$. Due to the constraint \eqref{eq:SnTe110}, Bloch wavefunctions on the (110) plane can be labeled by their mirror eigenvalue, such that the Hamiltonian becomes block-diagonal in the eigenbasis of $M_{xy}$. For $k_2=k_3\equiv k$, there exists a unitary matrix $U$ such that
\begin{equation}\label{eq:SnTeblockdiag}
 U^\dag H(k_1, k, k) U = \begin{pmatrix}
                          H_+ (k_1, k) & 0 \\
                          0 & H_- (k_1,k)
                         \end{pmatrix}
\end{equation}
takes the same form as \eqref{eq:afti_stack} on the mirror symmetric plane, where now $+$ and $-$ label the mirror eigenvalues, $m=\pm i$. In Appendix B we give a more detailed analysis of the atomic orbitals forming the $H_\pm$ layers in the basis Eq.~\eqref{eq:SnTeblockdiag}. The Chern numbers of the two-dimensional Hamiltonians $H_\pm$ can be evaluated using the method developed in Ref.~\onlinecite{Fukui2005}. We compute the integral of the Berry curvature, ${\boldsymbol\Omega}^\pm({\bf k}) = {\boldsymbol\nabla}_{\bf k} \times {\bf A}^\pm({\bf k})$, where
\begin{equation}\label{eq:berrycon}
 {\bf A}^\pm({\bf k}) = i \sum_n \langle \psi_n^\pm ({\bf k}) | {\boldsymbol\nabla}_{\bf k} | \psi_n^\pm ({\bf k}) \rangle
\end{equation}
is the Berry connection. Here, ${\bf k}=(k_1, k, k)$ is chosen to lie in the (110) plane, and $\psi_n^\pm$ is the $n^{\rm th}$ eigenstate of $H_\pm$, with mirror eigenvalue $m=\pm i$. Integrating over the Brillouin zone associated with the mirror invariant plane gives $C_+=-C_-=-2$, with $C_\pm=\int {\boldsymbol\Omega}^\pm({\bf k}) \cdot d{\bf S}$. The mirror Chern number equals $C_{(110)} = (C_{+} - C_{-})/2=-2$.

In the basis \eqref{eq:SnTeblockdiag} the system can be thought of as a stack of Chern insulators with alternating Chern number $\pm2$, similar to Fig.~\ref{fig:afti}b. Away from the $k_2=k_3$ plane the chiral edge modes of $H_+$ and $H_-$ are allowed to couple, leading to the formation of two surface Dirac cones pinned to the (110) plane of the surface BZ. 
The same analysis can be repeated for the (1$\overline{1}$0) plane, in which the protecting mirror symmetry maps $x\to -y$ and $y\to -x$. Here too we obtain $C_{(1\overline{1}0)}=-2$, confirming the topological origin of the second pair of surface Dirac cones. Therefore, in this parameter regime SnTe is topologically equivalent to two sets of layered Chern insulators, each having alternating invariants $C_\pm=\pm2$, which are stacked along the (110) and (1$\overline{1}$0) directions, respectively.

The same type of equivalence holds irrespective of the number of protected surface modes and their positions in the surface Brillouin zone. We highlight this fact by considering the effect of anisotropy in the Hamiltonian \eqref{eq:HSnTe}, which we model as a chemical potential imbalance of the $p_z$ orbitals,
\begin{equation}\label{eq:Hcrystalfield}
 H_{cf} = \sum_j (-1)^j \sum_{{\bf r},s} {\bf c}^\dag_{js}({\bf r})\cdot {\bf f}\, {\bf f}\cdot {\bf c}^\pd_{js}({\bf r}),
\end{equation}
with ${\bf f} = (0,0,f_z)$. Eq.~\eqref{eq:Hcrystalfield} describes the so called crystal field effect,\cite{Liu2015} which can appear as a consequence of the different chemical environments experienced by the $p_z$ and $p_{x,y}$ orbitals. For a SnTe monolayer in the (001) direction, this imbalance is given by the absence of $\sigma$-bonds between $p_z$ orbitals,\cite{Liu2015} and leads to a 2d TCI phase. In bulk SnTe, such a term may be generated by applying strain, or by intercalating topologically trivial spacer layers in the bulk crystal, thus forming a superlattice.

\begin{figure}[tb]
 \includegraphics[width=\columnwidth]{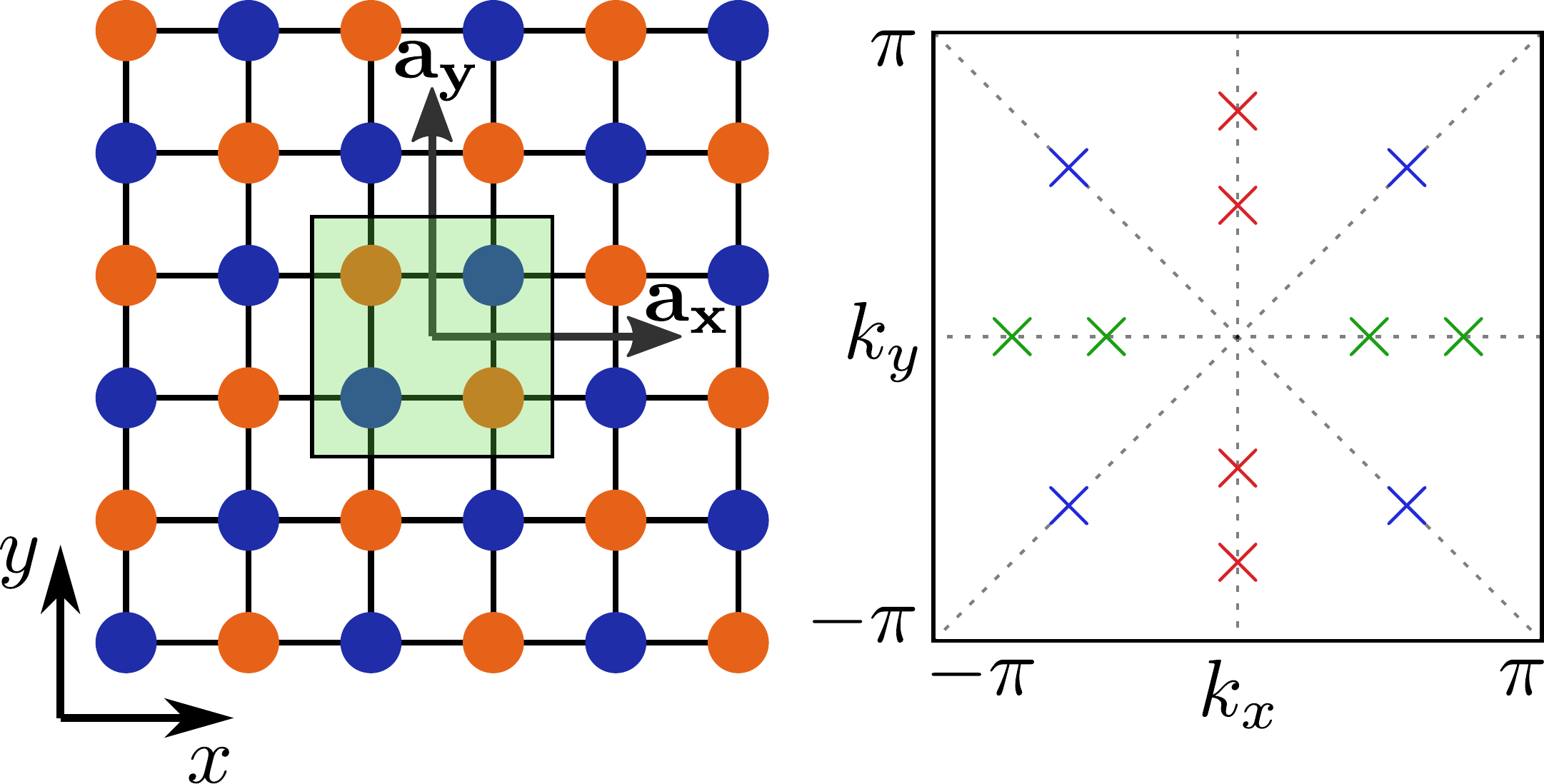}
 \caption{Left: top view of the surface perpendicular to the $z$-direction in an infinite slab geometry. The green box marks our choice of slab unit cell, with ${\bf a_x}$ and ${\bf a_y}$ the Bravais vectors. Right: corresponding surface BZ hosting 12 protected Dirac cones, marked with $\times$, located along the projections of the mirror symmetric planes (dashed lines). The blue Dirac cones are protected by (110) and (1$\overline{1}$0) mirror symmetries, while the red and green cones are protected by reflection symmetries about the Sn-Te planes, mapping $x\to-x$ and $y\to-y$, respectively.\label{fig:tci_12cones}}
\end{figure}

\begin{figure}[tb]
 \includegraphics[width=\columnwidth]{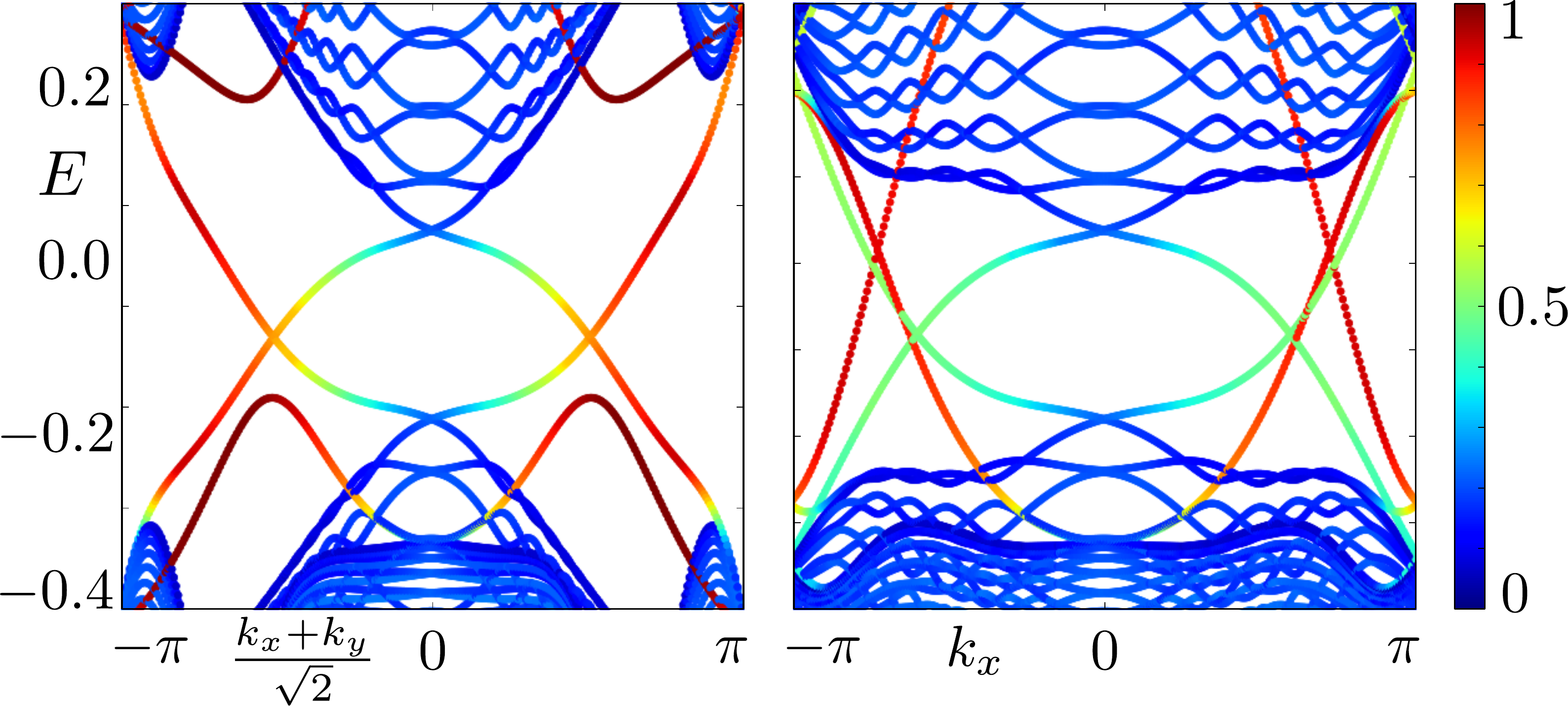}
 \caption{Bandstructure of the model \eqref{eq:HSnTe} in an infinite slab geometry with hard wall boundaries in the $z$ direction. We use a slab thickness of $80$ layers and the tight-binding parameters mentioned in the text. The color scale is given by the eigenstate intensity on the first and last $8$ layers from the boundaries. The left panel is a cut along the projection of the (110) plane on the surface BZ, showing a pair of surface Dirac cones. 
 The right panel is a cut along (100), on which there are two pairs of surface Dirac cones. The two pairs are distinguished by their different localization lengths, corresponding to red and green colors, respectively. \label{fig:tci_bands}}
\end{figure}

As the magnitude of the crystal field term is gradually increased, from $f_z=0$ to $f_z=-1$, the bulk gap closes and reopens, signaling a topological phase transition. We compute the bandstructure of the system for $f_z=-1$, using an infinite-slab geometry consisting of 80 monolayers, with hard-wall boundaries in the $z$-direction. Remarkably, the $E(k_x,k_y)$ dispersion shows a total of twelve surface Dirac cones in the presence of a crystal field, as shown in Figs.~\ref{fig:tci_12cones} and \ref{fig:tci_bands}.

The additional surface modes are also topologically protected, but by different lattice symmetries, namely reflections about one layer which map either $x\to -x$ or $y \to -y$. Both preserve the atomic species, taking a block diagonal form similar to \eqref{eq:Mxy}, but act differently on the $p$ orbitals and spin. Specifically,
\begin{equation}\label{eq:smallMx}
 {\cal M}_{x}    = \begin{pmatrix}
        -1 & 0 & 0 \\
        0 & 1 & 0 \\
        0 & 0 & 1
       \end{pmatrix}
\otimes i\sigma_x,
\end{equation}
and
\begin{equation}\label{eq:smallMy}
 {\cal M}_{y}    = \begin{pmatrix}
        1 & 0 & 0 \\
        0 & -1 & 0 \\
        0 & 0 & 1
       \end{pmatrix}
\otimes i\sigma_y.
\end{equation}

As before, on the (100) or (010) mirror-invariant planes the Hamiltonian becomes block-diagonal in the eigenbasis of Eqs.~\eqref{eq:smallMx} and \eqref{eq:smallMy}. We find mirror Chern numbers $C_{(100)}=C_{(010)}=4$, related by the four-fold rotation symmetry around the $z$ direction. 

While this need not be generically the case, in our specific tight binding model the addition of a crystal field leads to a tripling of surface Dirac cones: from four in the isotropic case to twelve with anisotropy. This proliferation is possible on a surface perpendicular to the $z$ direction because the term \eqref{eq:Hcrystalfield} only couples to the chemical potential of the $p_z$ orbitals, and therefore leaves all reflection symmetries acting in the $p_{x,y}$ space intact. Choosing a different field direction, by setting ${\bf f}=(f_x, 0, 0)$ in Eq.~\eqref{eq:Hcrystalfield} for instance, would break the (110) and (1$\overline{1}$0) mirror symmetries, and gap out the surface modes they protect.

When the surface does preserve the reflection symmetries, each symmetry can lead to surface Dirac cones, and the system is topologically equivalent to coupled Chern insulators stacked in the direction perpendicular to the reflection plane. This description remains valid also when protecting symmetries are not independent, as is the case in Fig.~\ref{fig:tci_12cones}. The two-dimensional surface Brillouin zone shows Dirac cones protected by a total of four reflection symmetries, around (110), (1$\overline{1}$0), (100), and (010), with the first two and the last two related to each other by a $C_4$ rotation symmetry, such that they have equal mirror Chern numbers.

\section{Conclusion}
\label{sec:conc}

We have shown how three-dimensional reflection symmetry protected TCIs can be obtained in layered structures consisting of Chern insulators with alternating topological invariants. The coupled-layer framework allows for an intuitive interpretation of surface Dirac cones, whose number and position in the surface BZ can be tuned. The Chern number of each layer determines the number of surface modes, the inter-layer coupling determines their position in the surface BZ, and the stacking direction controls which reflection symmetry is responsible for their protection. By combining systems that are layered in different directions, we can capture the simultaneous effect of multiple protecting symmetries. However, doing so results in TCIs that are no longer adiabatically connected to the limit of decoupled two-dimensional quantum Hall systems.

As an application, we have shown that surface modes in SnTe can be interpreted in the coupled-layer language. Using a simplified tight-binding model, we have found that anisotropy can lead to a substantial increase in the number of protected surface Dirac cones. Each set of surface modes can be seen as originating from an independent stack of Chern insulators, even when the protecting reflection symmetries are not independent from each other.

In this work, we have examined layers that host equal numbers of propagating edge modes with opposite chirality, in the presence of a coupling which only acts between neighboring layers.
More complex patterns, including longer range inter-layer coupling and unit cells composed of more than two layers may prove useful in describing other gapless topological phases protected by reflection symmetries.\cite{Chiu2014} While we have focused only on topological phases protected by reflection and their associated mirror Chern numbers, possible extensions of the coupled-layer approach to other spatial symmetries, such as glide symmetry, provide an interesting direction for future work.

The coupled-layer approach can be readily adapted to study particle-hole symmetric systems, by replacing the chiral electron modes in each layer with chiral Majorana modes.\cite{Qi2010} This can straightforwardly describe topological crystalline superconductors, and may provide a different tool-set for analyzing nodal superconductors protected by reflection symmetry.\cite{Chiu2014}

Finally, our work may provide an alternative way of incorporating the effect of electron-electron interactions in TCIs, which has been shown to modify their topological classification.\cite{Yoshida2015, Isobe2015} An interesting direction for future study would be to replace each of the layers with a fractional quantum Hall system, potentially leading to systems hosting fractional surface Dirac cones,\cite{Sagi2015} while still being able to control the number and position of gapless surface states. Recent progress in this direction has been reported in Ref.~\onlinecite{Song2016}.

\acknowledgments

NA and HB acknowledge support from the European Research Council (\#678702 ``TOPO-NW'') and the Israeli Science Foundation. ICF and AS thank the European Research Council under the European Union`s Seventh Framework Programme (FP7/2007-2013) / ERC Project MUNATOP, the US Israel Binational Science Foundation, and the Minerva Foundation for support.

\bibliography{tci}

\end{document}


\title{Supplemental material to: ``Coupled-layer description of topological crystalline insulators''}

\author{I. C. Fulga}
\author{N. Avraham}
\author{H. Beidenkopf}
\author{A. Stern}
\affiliation{Department of Condensed Matter Physics, Weizmann Institute of Science, Rehovot 76100, Israel}

\date{\today}

\maketitle

\appendix

\section{Multiple mirror symmetries}
\label{app:rxrz}

In this Section we give an example of a concrete tight-binding model realizing a TCI protected by two reflection symmetries. We consider a system on a cubic lattice, composed of two sets of coupled Chern insulators with alternating Chern numbers, stacked in the $x$ and $z$ directions, respectively. The momentum space Hamiltonian reads
\begin{widetext}
\begin{equation}\label{eq:apphrxrz}
 H = \begin{pmatrix}
      H_l(k_x,k_y) & T_z \sin(k_z) & 0  & 0  \\
      T_z \sin(k_z) & H_l(-k_x,k_y) & 0  & 0  \\
      0  & 0  & H_l(k_z,k_y) & T_x \sin(k_x) \\
      0  & 0  & T_x \sin(k_x) & H_l(-k_z,k_y)
     \end{pmatrix},
\end{equation}
\end{widetext}
where $H_l$ is the Hamiltonian describing a single layer. For the latter, we choose a two-band model for a Chern insulator
\begin{equation}\label{eq:applayer}
\begin{split}
H_l(k_1,k_2) = & [\mu - 2\cos(k_1) - 2\cos(k_2)]\sigma_z \\
& + \sin(k_1)\sigma_x+\sin(k_2)\sigma_y,
\end{split}
\end{equation}
where $\sigma_i$ are Pauli matrices. The inter-layer coupling terms are chosen to be equal, taking the form $T_x=T_z=t\sigma_0$.

The full system obeys two reflection symmetries with anti-commuting operators
\begin{equation}\label{eq:apprxmat}
 R_x = \begin{pmatrix}
        0 & 1 & 0 & 0 \\
        1 & 0 & 0 & 0 \\
        0 & 0 & 1 & 0 \\
        0 & 0 & 0 & -1 \\
       \end{pmatrix},
\end{equation}
and
\begin{equation}\label{eq:apprzmat}
 R_z = \begin{pmatrix}
        1 & 0 & 0 & 0 \\
        0 & -1 & 0 & 0 \\
        0 & 0 & 0 & 1 \\
        0 & 0 & 1 & 0 \\
       \end{pmatrix},
\end{equation}
such that
\begin{equation}\label{eq:apprefl_x}
 R^\pd_x H(k_x,k_y,k_z) R_x^{-1} = H(-k_x,k_y,k_z)
\end{equation}
and
\begin{equation}\label{eq:apprefl_z}
 R^\pd_z H(k_x,k_y,k_z) R_z^{-1} = H(k_x,k_y,-k_z).
\end{equation}

\begin{figure}[tb]
 \includegraphics[width=\columnwidth]{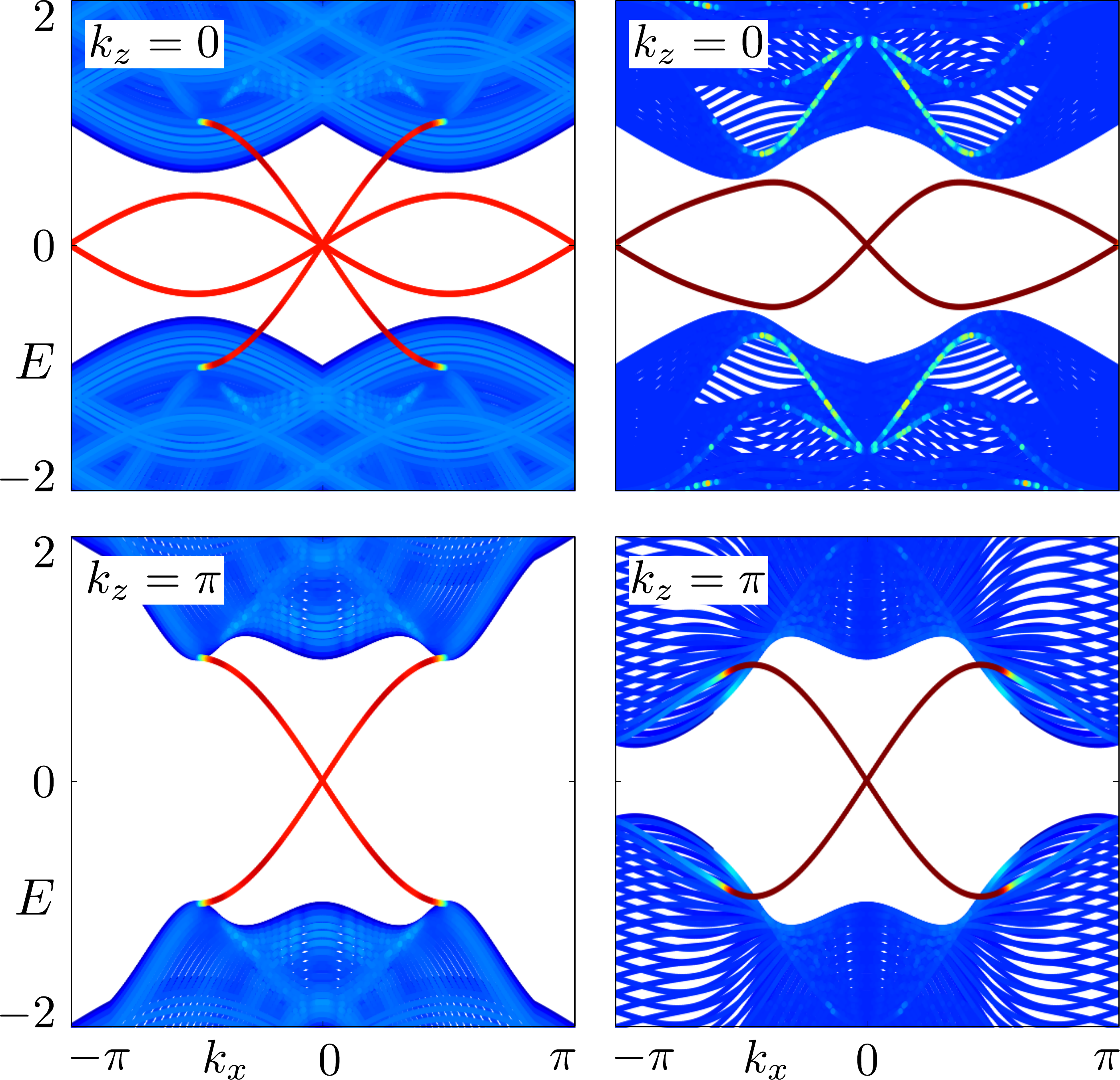}
 \caption{Bandstructures of the $R_x$ and $R_z$ symmetric system in an infinite slab geometry (infinite along the $x$ and $z$ directions, 40 unit cells wide along the $y$ direction), for fixed $k_z=0,\pi$, as a function of $k_x$. Bulk states are shown in blue, while surface modes are plotted in red. Only surface modes appearing on the $y=0$ surface are shown. Left panels: the Hamiltonian Eq.~\eqref{eq:apphrxrz} has mirror Chern numbers $(C_{k_x,0}, C_{k_x,\pi}, C_{k_z,0}, C_{k_z,\pi})=(1,1,-1,-1)$. Right panels: adding the terms Eq.~\eqref{eq:appmuxz} and \eqref{eq:appoffdiagh} leads to mirror Chern numbers $(C_{k_x,0}, C_{k_x,\pi}, C_{k_z,0}, C_{k_z,\pi})=(1,0,-1,0)$. Nonzero values of these invariants lead to surface Dirac cones which connect the valence and conduction bands. Bandstructures at fixed $k_x=0,\pi$ as a function of $k_z$ are identical, due to a $x \leftrightarrow z$ reflection symmetry.\label{fig:appbands}}
\end{figure}

Setting $\mu=2$ and $t=0.4$, we obtain a TCI in which the four mirror Chern numbers ($C_{k_x,0}$, $C_{k_x,\pi}$, $C_{k_z,0}$, $C_{k_z,\pi}$) take the values $(1, 1, -1, -1)$. For a system which is finite in the $y$ direction, the $x - z$ BZ shows a total of four surface Dirac cones (see Fig.~\ref{fig:appbands}, left panels). Two are protected by the $R_x$ mirror symmetry and located at $(k_x,k_z)=(0,0)$ and $(\pi, 0)$, while the other two are due to the $R_z$ reflection symmetry and positioned at $(k_x,k_z)=(0,0)$ and $(0, \pi)$.

To change the values of the four invariants while respecting the mirror symmetries Eq.~\eqref{eq:apprefl_x} and \eqref{eq:apprefl_z}, the system must undergo a topological phase transition, signaled by a closing and reopening of the bulk gap. We add a momentum-dependent mass term to each of the layers, replacing
\begin{equation}\label{eq:appmuxz}
H_l\to H_l + m[1+\cos(k_x)][1+\cos(k_z)]\sigma_z
\end{equation}
with a value of $m=0.5$, which leads to a bulk gap closing at the $\Gamma$-point $(k_x,k_z)=(0,0)$ while leaving the bandstructures at $k_x=\pi$ and $k_z=\pi$ unaffected. 
To reopen the bulk gap, we add off-diagonal terms to the Hamiltonian Eq.~\eqref{eq:apphrxrz}, which read
\begin{equation}\label{eq:appoffdiagh}
\Lambda = \begin{pmatrix}
1 & 1 \\
1 & -1
\end{pmatrix}\otimes  a [\cos(k_x)+\cos(k_z)]\sigma_z.
\end{equation}

Both the terms Eq.~\eqref{eq:appmuxz} and \eqref{eq:appoffdiagh} respect the $R_x$ and $R_z$ mirror symmetries. Setting a value $a=-0.6$ leads to a system with a fully gapped 3d bulk, but with mirror Chern numbers  $(C_{k_x,0}, C_{k_x,\pi}, C_{k_z,0}, C_{k_z,\pi})=(1,0,-1,0)$ (see Fig.~\ref{fig:appbands}, right panels). Note that Eq.~\eqref{eq:appmuxz} corresponds to an inter-layer hopping, which connects layers with the same Chern number, both for the stack oriented in the $x$ direction and for the one oriented in the $z$ direction. Unlike the examples considered in the main text, here the coupling term Eq.~\eqref{eq:appmuxz} cannot be made arbitrarily weak, but requires a value of $m$ large enough to close and reopen the bulk gap. In this case the layer construction does not adiabatically connect the phase to a stack of decoupled layers, but does provide an intuitive way to construct the TCI from a set of mirror-symmetric layers.

\section{Coupled layers in SnTe}
\label{app:snte}

In this Section we discuss the basis change which renders the SnTe tight-binding model of the main text block-diagonal along the $(110)$ mirror invariant plane. The momentum space Hamiltonian is a $12\times12$ matrix encoding the degrees of freedom associated to the Sn and Te atomic species, each of which is modeled as having three $p$-orbitals ($p_x$, $p_y$, $p_z$) and two spin orientations ($\uparrow$, $\downarrow$). The $(110)$ mirror symmetry of Eqs.~$(18-20)$ in the main text is written in the basis ($|{\rm Sn},p_x,\uparrow\rangle$,$|{\rm Sn},p_x,\downarrow\rangle$,$|{{\rm Sn},p_y,\uparrow\rangle}$,
${|{\rm Sn},p_y,\downarrow\rangle}$,${|{\rm Sn},p_z,\uparrow\rangle}$,$|{\rm Sn},p_z,\downarrow\rangle$,$|{\rm Te},p_x,\uparrow\rangle$,$|{\rm Te},p_x,\downarrow\rangle$,\\$|{\rm Te},p_y,\uparrow\rangle$,$|{\rm Te},p_y,\downarrow\rangle$,$|{\rm Te},p_z,\uparrow\rangle$,$|{\rm Te},p_z,\downarrow\rangle)$. After the basis change of Eq.~$(21)$ in the main text, the Hamiltonian becomes block diagonal, corresponding to two dimensional layers $H_\pm$ with opposite Chern numbers. The states characterizing the $H_\pm$ blocks, $|\psi_\pm\rangle$, are linear combinations of the 12 orbitals:
\begin{equation}\label{eq:apppsiplus}
|\psi_+\rangle=\frac{1}{2}\begin{pmatrix}
-(1+i)|{\rm Sn},p_x,\downarrow\rangle-\sqrt{2}|{\rm Sn},p_y,\uparrow\rangle \\
 -(1+i)|{\rm Sn},p_y,\downarrow\rangle-\sqrt{2}|{\rm Sn},p_x,\uparrow\rangle \\
 -(1+i)|{\rm Sn},p_z,\downarrow\rangle-\sqrt{2}|{\rm Sn},p_z,\uparrow\rangle \\
 -(1+i)|{\rm Te},p_x,\downarrow\rangle-\sqrt{2}|{\rm Te},p_y,\uparrow\rangle \\
 -(1+i)|{\rm Te},p_y,\downarrow\rangle-\sqrt{2}|{\rm Te},p_x,\uparrow\rangle \\
 -(1+i)|{\rm Te},p_z,\downarrow\rangle-\sqrt{2}|{\rm Te},p_z,\uparrow\rangle \\
\end{pmatrix}
\end{equation}
and
\begin{equation}\label{eq:apppsimin}
|\psi_-\rangle=\frac{1}{2}\begin{pmatrix}
(1+i)|{\rm Sn},p_x,\downarrow\rangle-\sqrt{2}|{\rm Sn},p_y,\uparrow\rangle \\
 (1+i)|{\rm Sn},p_y,\downarrow\rangle-\sqrt{2}|{\rm Sn},p_x,\uparrow\rangle \\
 -(1+i)|{\rm Sn},p_z,\downarrow\rangle+\sqrt{2}|{\rm Sn},p_z,\uparrow\rangle \\
 (1+i)|{\rm Te},p_x,\downarrow\rangle-\sqrt{2}|{\rm Te},p_y,\uparrow\rangle \\
 (1+i)|{\rm Te},p_y,\downarrow\rangle-\sqrt{2}|{\rm Te},p_x,\uparrow\rangle \\
 -(1+i)|{\rm Te},p_z,\downarrow\rangle+\sqrt{2}|{\rm Te},p_z,\uparrow\rangle \\
\end{pmatrix}.
\end{equation}
The $H_\pm$ layers lie on top of each other in real space, and are composed of both the Sn and Te atomic orbitals.